\documentclass[aps,twocolumn,showpacs,preprintnumbers]{revtex4}
\usepackage{graphicx,amsmath,hhline} 
\usepackage{newcent}
\begin{document}
\title{Dynamics of social networks}
\author{Holger Ebel}
\email{ebel@theo-physik.uni-kiel.de}
\affiliation{Institut f\"ur Theoretische Physik, Universit\"at Kiel,
Leibnizstr.\ 15, D-24098 Kiel, Germany}
\author{J\"{o}rn Davidsen}
\thanks{Present address: Max-Planck-Institute for Physics of Complex Systems, N\"{o}thnitzer Str.\ 38, D-01187 Dresden, Germany.\\ Electronic address: davidsen@mpipks-dresden.mpg.de}
\affiliation{Chemical Physics Theory Group, Department of Chemistry, University of Toronto, Toronto, Canada M5S 1A1}
\author{Stefan Bornholdt}
\email{bornholdt@izbi.uni-leipzig.de}
\affiliation{Interdisziplin\"ares Zentrum f\"ur Bioinformatik, Universit\"at Leipzig, Kreuzstr.\ 7b, D-04103 Leipzig, Germany}

\begin{abstract}
Complex networks as the World Wide Web, the web of human sexual contacts or criminal networks often do not have an engineered architecture but instead are self-organized by the actions of a large number of individuals. From these local interactions non-trivial global phenomena can emerge as small-world properties or scale-free degree distributions. A simple model for the evolution of acquaintance networks highlights the essential dynamical ingredients necessary to obtain such complex network structures. The model generates highly clustered networks with small average path lengths and scale-free as well as exponential degree distributions. It compares well with experimental data of social networks, as for example coauthorship networks in high energy physics.
\end{abstract}
\pacs{89.75.Hc,87.23.Ge,89.75.Da}
\preprint{{\em Complexity} (2003), accepted for publication (Proceedings of {\em Concepts for Complex Adaptive Systems} 2002)}
\maketitle

\section{Introduction}
In many kinds of complex systems large and stable network structures occur. Specific examples include networks of interacting proteins or genes, ecological graphs, communication networks, and social networks \cite{strogatz:2001,albert/barabasi:2002,dorogovtsev/mendes:2002,sporns/tononi:2002}. For most of them, neither random networks nor regular lattices provide an adequate framework to model the observed topological properties. The first step towards an improved understanding was the mathematical concept of ``small-world networks" introduced by Watts and Strogatz \cite{watts/strogatz:1998,watts:1998}. Small-world networks interpolate between the two limiting cases of a highly clustered regular lattice and a random graph with short path lengths between nodes. A network is said to be highly clustered in the sense that if node A is linked to node B and B is linked to node C, there is an enhanced probability that A will also be linked to C (a property that sociologists call ``transitivity"). The distance between two nodes is defined as the number of edges along the shortest path connecting them. If a network shares the following two characteristic properties it is called a ``small-world" \cite{watts/strogatz:1998,amaral/stanley:2000}: (i) high clustering and (ii) a small average shortest path between two nodes scaling logarithmically with network size.
Social networks and acquaintance networks in particular are typical examples for small-world behavior \cite{strogatz:2001,albert/barabasi:2002,dorogovtsev/mendes:2002,amaral/stanley:2000}. The topology and the static characteristics of them have been the focus of recent investigations as, for example, in the case of illegal and terrorist networks \cite{baker/faulkner:1993,krebs:2002} or the web of human sexual contacts \cite{liljeros/aberg:2001}. It is quite dissatisfying that much less is known about the dynamical properties of these systems because such real-world networks generally are not static but evolve in time. Thus, the emergence of networks displaying small-world behavior should be directly related to local interactions within the network.
Besides clustering and path lengths, the degree distribution further characterizes a complex network. Of particular interest are scale-free networks where the degree (i.e., the number of a node's next neighbors) is distributed according to a power law. Such scale-free degree statistics lead to distinct behavior with respect to error and attack tolerance \cite{albert/barabasi:2000b} or epidemic spreading \cite{pastor-satorras/vespignani:2001} and is observed in some social networks \cite{strogatz:2001,albert/barabasi:2002,dorogovtsev/mendes:2002,newman:2001b,barabasi/vicsek:2001,ebel/bornholdt:2002}. In many cases, the origin of scale-free properties is well understood in terms of interactions that generate this topology dynamically, e.g.\ on the basis of network growth and preferential linking \cite{albert/barabasi:2002,dorogovtsev/mendes:2002,bornholdt/ebel:2001}. While these models generate scale-free structures they do not, in general, lead to clustering and are therefore of limited use when modeling small-world networks and social networks in particular.
Here, dynamics of social networks and the emergence of a small-world structure is addressed by combining ideas from the two fields of ``small-world networks" and ``scale-free networks" \cite{davidsen/bornholdt:2002}. More precisely, starting with the example of a coauthorship network, a simple dynamical model is introduced generating highly clustered networks with small average path lengths that scale logarithmically with network size. In addition to its small-world behavior, this model leads to a scale-free degree distribution for small death-and-birth rates of nodes.

\section{Social webs: a coauthorship network}
Coauthorship networks are well-studied examples of social networks reflecting the emergence of cooperative structures between scientists. Nodes of such networks are authors which are connected if they have coauthored a paper together. One example is the coauthorship network of  high energy physicists which was reconstructed by Newman from the SPIRES publication database for the years 1995-1999 \cite{newman:2001b,newman:2001e}. It contains $55\,627$ authors as nodes with a mean degree of $\langle k_S \rangle = 173 $ and an average shortest path length of $\ell_S   = 4.0$. Since the degree distribution is consistent with a power law
$P(k) \propto k^{- \gamma}
$
 with exponent  $\gamma  = 1.2$, this coauthorship network belongs to the class of scale-free social networks. The clustering of the network is measured by the clustering coefficient $C$ defined as follows \cite{watts/strogatz:1998}. First the density of links in the neighborhood of an individual node $i$ is given by the ratio of existing links $E_i$ to the potential number of connections $1/2\, k_i \,(k_i -1)$. Then the clustering coefficient of the entire network is the average density
\begin{equation}
C = \left\langle \frac{2 E_i}{k_i (k_i -1)} \right\rangle.\label{def_c}
\end{equation}
A similar, but not equivalent, definition for the clustering coefficient is provided by the fraction of fully connected ``triples" with a triple being a connected subgraph containing three nodes \cite{newman/watts:2001,barrat/weigt:2000}
\begin{equation}
C_{\Delta} = \frac{3 \times (\text{number of fully connected triples})}{
\text{number of triples}}.\label{def_cd}
\end{equation}
The latter definition is equivalent to reversing the order of averaging and division in (\ref{def_c}). We computed the clustering coefficient from the raw data using definition (\ref{def_c}) to $C_S  = 0.68$, whereas $C_{\Delta, S}   = 0.73$ was calculated for definition (\ref{def_cd}) \cite{newman:2001b}. To understand the meaning of these values of clustering and path length, we will compare them to the respective quantities of random networks of identical size. A naive approach would be to consider a random network with identical mean degree where each pair of nodes is connected with a constant probability \cite{davidsen/bornholdt:2002}, yielding $C_{\text{rand}}  = 0.0031$ and $\ell_{\text{rand}}  = 2.12$. However, comparison with this particular type of random network is flawed since the constant linking probability results in a Poissonian degree distribution that differs strongly from the observed scale-free behavior. Therefore, we deduce both quantities for a random network with identical degree distribution but randomly assigned links. With the estimate for the clustering coefficient according to definition (\ref{def_c})  \cite{davidsen/bornholdt:2002}
\begin{equation}
C' = \frac{1}{\langle k \rangle N} \biggl( \frac{\langle k^2 \rangle}{\langle k \rangle} -1 \biggr),\label{eq_c}
\end{equation}
a value $C'  = 0.12$ is obtained, much larger than $C_{\text{rand}}$  but still smaller than the observed clustering. Thus, the network exhibits very high clustering which is not explained by the degree distribution alone. Note that the estimate (\ref{eq_c}) holds exactly in the case of a Poissonian degree distribution and, in the thermodynamic limit, for definition (\ref{def_cd}), too \cite{newman:2002b}. The path length in a random network with the same degree distribution  \cite{davidsen/bornholdt:2002}, $\ell'   = 1.81$, is even smaller than in a random network with constant linking probability. This is caused by the highly connected hubs present in a scale-free network. That  $\ell'$ is slightly smaller than the observed value is due to the fact that many links are consumed for building the densely connected neighborhoods. Hence, the SPIRES coauthorship network exhibits pronounced small-world behavior and, in addition, shows scale-free behavior in terms of the link distribution.

\section{Modeling social networks}
\begin{figure}
\includegraphics[width=5.5cm,angle=-90]{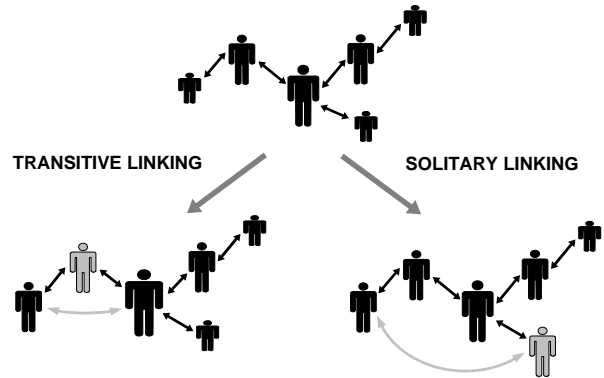}
\caption{\label{fig_linking}The basic mechanism of the network model. A new link is formed by a randomly chosen individual (gray) introducing two of its acquaintances, that have not met before, to each another ({\em transitive linking}). If the chosen person has less than two acquaintances he introduces himself to another person which he picks out at random ({\em solitary linking}). The size of the figures corresponds to their respective degree.}
\end{figure}
Let us now consider a model for social networks in terms of acquaintance graphs with persons as nodes and undirected links between people who know each other  \cite{davidsen/bornholdt:2002}. The acquaintance network evolves with new acquaintances forming between individuals, and people joining and leaving the network. We assume that the central mechanism of the dynamics of acquaintance networks is that people are introduced to each other by a common acquaintance ({\em transitive linking}). The dynamics of the model consist of two processes taking place at each time step:
(i) One individual is chosen at random and introduces two arbitrary acquaintances to each other. They become acquainted to each other, too, if they have not met before, and a new link is build ({\em transitive linking}). If the chosen person has less than two acquaintances he will introduce himself to an individual picked out at random (Fig.\ \ref{fig_linking}).
(ii) With probability $p$, a randomly chosen person leaves the network and all the links between him and his acquaintances are deleted. A new individual joins the network and becomes acquainted to one randomly chosen person.
Note that the number of nodes $N$ remains constant, neglecting fluctuations in the number of individuals being part of the acquaintance network. The finite lifetime of links leads to a stationary state of the network approximating the behavior of real social networks. However, it is in contrast to most dynamical network models which are based on network growth \cite{albert/barabasi:2002,dorogovtsev/mendes:2002,barabasi/vicsek:2001,klemm/eguiluz:2002}. The two time scales of the model are separated by the probability $p$. The rate of building new social connections can be as short as minutes or hours, whereas the time scale of joining and leaving the network may lie in the range of years or decades. Hence, in the following, we will focus on small death-and-birth rates, $p \ll 1$.

\section{Degree distribution, path length, and clustering}
\begin{figure}
\includegraphics[width=8cm,height=6.5cm]{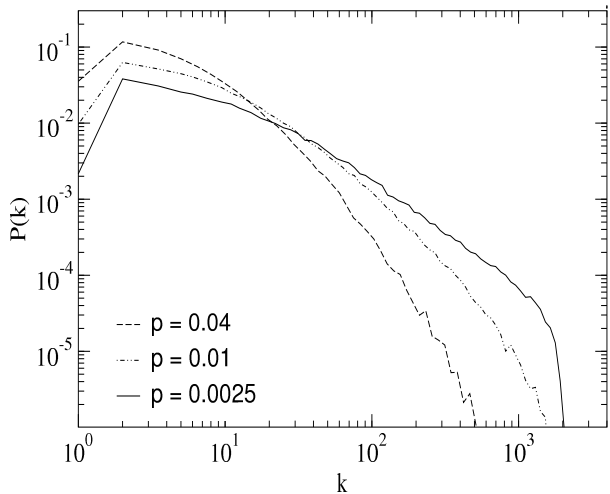}
\caption{\label{fig_pk}Degree distributions $P(k)$ for different values of $p$. The degree distribution shows power-law behavior for small $p$ with an exponent of $-1.35$ ($p = 0.0025$, $N = 7000$). The distribution is largely insensitive to system size $N$ with the cutoff being caused by the finite age of nodes.}
\end{figure}
Once the network has reached a stationary state, one can characterize the resulting network by its degree distribution $P(k)$. Results of numerical simulations for different values of $p$ are shown in Fig.\ \ref{fig_pk}. The number of acquaintances a node can collect are limited by its finite lifetime corresponding to the cutoff of the degree distribution at high $k$. With $p\ll 1$, the dynamics are dominated by the transitive linking process (i) giving rise to a power-law decay over a certain range which increases with decreasing $p$. For larger values of $p$, the Poissonian death-and-birth process (ii) competes with the transitive linking (i) which leads to a stretched exponential range in the distribution until the Poissonian dynamics of (ii) dominates. Depending on the death-and-birth rate $p$, the above model is able to generate degree distributions covering scale-free and exponential regimes all being observed in real-world networks. For sufficiently large graphs, these distributions solely depend on $p$, the single free parameter of the model. Experimental data suggest low values of the death-and-birth rate, $p \ll 1$, such that the two time scales of network dynamics are well separated.
\begin{table}
\begin{tabular*}{8.5cm}{@{\extracolsep{\fill}}|ccccc|}
\hline
$p$ & $\langle k \rangle$ & 
$C$ & $C'$ & $C_{\text{rand}}$ \\
\hline
0.04 & 14.9 & 0.45& 0.036 & 0.0021 \\
0.01 & 49.1  & 0.52 & 0.29 & 0.0070\\
0.0025 & 149.2 & 0.63 & 0.43 & 0.021 \\
\hline
\end{tabular*}
\caption{\label{tab}Clustering coefficient $C$ ($p$: birth-and-death rate, $N = 7000$). $C'$ is an estimate for the average clustering coefficient of a network with identical degree distribution $P(k)$, but without transitive linking. 
$C_{\text{rand}}$ is the clustering coefficient of a random network with same size and constant linking probability.}
\end{table}
The average shortest path length $\ell$ is calculated directly from the stationary networks and shows a logarithmic scaling with system size \cite{davidsen/bornholdt:2002}. Using the data of Table \ref{tab}, the values $\ell'  = 1.59$ and $\ell_{\text{rand}}  = 1.77$ are calculated for $p = 0.0025$. Similarly, the average shortest path length of our model yields the very low value of $\ell = 2.38$. This, together with the logarithmic scaling of $\ell$, verifies that networks evolved by the simple rule of transitive linking meet the first requirement of small-world behavior.
Applying definition (\ref{def_c}), the clustering coefficient $C$ can be easily related to the mean degree $\langle k \rangle$  and the birth-and-death rate $p$  \cite{davidsen/bornholdt:2002}
\begin{equation}
1 - C = p \left( \langle k \rangle - 1 \right).\label{eq_cpk}
\end{equation}
As can be calculated from Table \ref{tab}, Eq.\ (\ref{eq_cpk}) yields the same values of the clustering coefficient $C$ which are obtained numerically for different values of $p$. As required for the second small-world property, the clustering is far higher than for a random network. Moreover, the observed clustering $C$ is not as strongly dependent on the mean degree as the respective values $C'$ and $C_{\text{rand}}$ , which is directly explained by Eq.\ (\ref{eq_cpk}). Altogether, this results in pronounced small-world behavior of the presented model.

\section{Discussion and conclusions}
The example of the SPIRES coauthorship network demonstrates how our model can be applied to a social network in the dynamically stationary state. For the regime of small turnover rates $p \ll 1$, small-world properties and scale-free behavior of the coauthorship network are reproduced by our model. Applying the logarithmic scaling for $l$, the average shortest path length of the model is in accordance with the experimental value. The clustering coefficients agree, too, and the model exhibits a scale-free degree distribution similar to the real-world network. Furthermore, this model provides a suitable framework for the study of other small-world networks, with exponential or broad degree distributions in particular. This is further confirmed by the fact that the results presented here are not altered when a small amount of noise is added to the link forming process. 
In conclusion, the model presented here provides a framework for the study of complex social networks interpolating between networks with scale-free and exponential degree distributions. The small-world properties of high clustering and small mean shortest path length, observed in may real-world systems, are achieved by the local rule of transitive linking. The statistical and topological properties of the network depend on the single free parameter of the model, the turnover rate $p$, which can be related to the rate of nodes entering and leaving the network in the stationary state. Complex networks are sometimes viewed as the ``backbone" of a complex system. Revealing the basic building mechanism for a broad class of social networks adds to the understanding how these skeletons can emerge from local dynamical rules.

\begin{acknowledgements}
We thank Mark Newman for sharing his data of the coauthorship network with us.
\end{acknowledgements}

\end{document}